\documentclass[nolongbibliography,prd,twocolumn,nofootinbib,showkeys,showpacs,preprintnumbers,amsmath,amssymb]{revtex4-1}

\usepackage{epsfig}
\usepackage{graphicx}
\usepackage{dcolumn}
\usepackage{graphics}
\usepackage{bm}
\usepackage{subfigure}
\usepackage{color}
\usepackage[normalem]{ulem}
\usepackage{soul}

\usepackage{bbold}

\begin{document}

\preprint{APS/123-QED}

\title{Anomalous Cyclic in the Neutrino Oscillations}

\author{E. Aydiner}
\email{ekrem.aydiner@istanbul.edu.tr}

\affiliation{Department of Physics, Faculty of Science, Istanbul University 34134, Istanbul Turkey}

\date{12 February 2022, Moda}

\begin{abstract}
Neutrino physics is one of the most important topics in particle physics and cosmology. Despite the many physical properties of neutrinos that are understood theoretically and experimentally, it is known that there are many unsolved problems in this area. In this study, we suppose that the deformed space-time caused by the gravitational perturbation can play an important role in neutrino oscillations. We analytically analyzed these effects on the neutrino oscillation and showed that this effect leads to an anomalous cyclic in the neutrino oscillation. The results clearly indicate that these anomalous cyclics depend on the degree of deformation of space-time. The role of this anomaly in neutrino oscillation may be important. There may be a relation between cyclic anomaly and other anomalies of neutrinos such as mass limits, energy limits, oscillation lengths, mixing angles, and speed of neutrinos. The cyclic anomalies might be detected  experimentally. If these anomalies are confirmed, it will appear that we need to think more about neutrino physics.
\end{abstract}


\maketitle



\section{Introduction} 

In the 1950s Pontecorvo proposed the massive neutrinos for the first time. Afterward, in 1957 he suggested a practical method for investigating neutrino oscillations in the vacuum setting an analogy with kaon oscillations \cite{Pontecorvo1958,Pontecorvo1968,Gribov1969}.  In  Pontecorvo’s scenario, the mixing between the flavor and mass eigenstates of neutrinos leads to oscillation. This model imposes that the neutrinos have a tiny mass unlike the standard model of particle physics. As we will show later, each type of oscillation is completely periodic, although the periods of neutrino oscillations vary depending on the source. The theoretically proposed neutrino oscillation has also been observed experimentally \cite{Kajita1998,Ahmad2001}.

In this study, we propose for the first time that anomalous cycles can occur in neutrino oscillations. Before going into the theory of neutrino oscillation and anomalous cycle, we will briefly introduce the historical background of neutrinos.

Historically, the neutrino $\nu$ was proposed in December 1930 by Pauli in order to explain the continuous energy-spectrum of the electrons measured in $\beta$-decays. Pauli named these particles neutrons because of their uncharged nature. However, after the discovery of the heavier and uncharged particle by Chadwick in 1932, Fermi changed this name to the neutrino.

According to our knowledge that the neutrinos are left-handed fundamental particles like quarks, photons and electrons 
and they are generated by weak interactions $W^{\pm}$and $Z^{0}$ which are carriers of the force between fermions. They cannot decay into other particles and do not carry an electric charge. Therefore, it is very difficult to detect neutrinos experimentally. However, using unusual experimental setups, in 1956, the electron neutrino $\nu_{ e}$ in the inverse $\beta$ decay was firstly discovered by Reines and Cowan \cite{Cowan1956}. This pioneering discovery was awarded the Nobel Physics Prize in 1995. In 1962, the muon neutrino $\nu_{ \mu}$ interactions were detected by Lederman, Schwartz, and Steinberger \cite{Lederman1962}. Lederman, Schwartz, and Steinberger shared Nobel Prize in 1988 due to the neutrino beam method and the demonstration of the doublet structure of the leptons through the discovery of the muon neutrino.  Finally,  the tau particle was detected in a series of experiments between 1974 and 1977 by Perl with his colleagues at the SLAC–LBLgroup \cite{Perl1975}.  Perl shared the 1995 Nobel Prize in Physics with Reines for the discovery of the tau neutrino. Additionally, in 2000, the interactions of tau neutrino $\nu_{ \tau}$  were firstly confirmed by the DONUT collaboration at Fermilab \cite{Kodama2001}.

In the 1960s, the solar neutrino problem remained unsolved for nearly three decades. However, in order to explain this inconsistency, it has been suggested that neutrinos may have mass and oscillate between flavors. In the following years, the existence of solar and cosmic neutrinos was investigated. In the 1960s, the flux of electron neutrinos coming from sun in the Homestake experiment and a discrepancy between the results of the experiment and the predictions of the Standard Model was found \cite{Davis1968}. This discrepancy was called as the solar neutrino problem. Furthermore, Koshiba also studied solar and cosmic neutrino detection. Particularly, he observed neutrinos from the SN 1987A supernova in the nearby Large Magellanic Cloud \cite{Hirata1987,Bionta1987}.  Therefore, Davis and Koshiba shared Nobel Physics Prize in 2002 for the detection of sun and cosmic neutrinos. 

To understand the solar neutrino problem and confirm the neutrino oscillation, many experimental studies have been realized such as in Sudbury Neutrino Observatory (SNO) and Super-Kamiokande. In 1998, it was detected at the Super-Kamiokande neutrino detector that neutrinos have a tiny mass and they oscillate from one flavor to another \cite{Kajita1998}. On the other hand, the first experimental evidence of neutrino oscillation was published in 2001 by SNO \cite{Ahmad2001}. Experimental findings revealed that neutrinos coming from the sun oscillate. These results also supported that neutrinos would have mass. Kajita and McDonald received the 2015 Nobel Prize for Physics.  

As a result, it is experimentally shown that they have  a tiny mass compared to other elementary particles \cite{Kajita1998,Ahmad2001,Kajita2015,Agafonova2010,Mertens2016,Olive2016}. Neutrino masses are so small that so far no experiment has succeeded in measuring them. It is assumed that the masses of all fundamental particles come from the Higgs field, but neutrino might get their masses another way. Since the mass of each neutrino is unknown, the upper bound on the total mass of the three neutrinos is obtained from cosmological calculations. Accordingly, the total mass of the three neutrinos should not exceed 50 eV \cite{Hut1979}.  Some research results show that the mass of the neutrino should be about one million times less than the mass of the electron \cite{Mertens2016,Olive2016}.

Today, we know that neutrinos are created by many processes in nature such as artificial nuclear reactions, nuclear reactions in stars, particle decay processes, explosions of the supernova and the spin-down of a neutron stars. However, these particles are the most abundant in Universe after the photon, therefore, this indicates that the main source must be the cosmic beginning of the universe. Moreover, we know that neutrinos are well-known six in number with anti-neutrinos. 

It should be stated that the physical properties of the neutrino have been experimentally investigated by using many defectors such as  IMB, MACRO, Kamiokande II, Super-Kamiokande, SNO, LSND, MINOS, DONUT, CERN, OPERA, MiniBooNE, Fermilab, DUNE and KATRIN. Despite the significant achievements of neutrino physics, one can see that there are many open problems in this area that arise from theory and experiments. For instance, although neutrinos have tiny masses, however, the absolute mass scale is still not known. It is known that neutrino oscillations are sensitive only to the difference in the squares of the masses \cite{Mohapatra2007,Amsler2008}. Neutrino masses were measured by different experiments however, the question of how neutrino masses arise has not been answered conclusively.

On the other hand, in the standard model, left- and right-handed versions of the fermion can be modeled. Unfortunately, right handed neutrinos have not been observed in the experiments. To solve this problem, the most conventional way is to add the right-handed neutrinos with very large Majorana masses to the standard Lagrangian. This is known as the seesaw mechanism which denotes beyond of standard model. Theoretically, the seesaw mechanism requires the existence of heavy $\nu_{ R}$’s or other appropriate beyond Standard Model physics at very high energies, which is called sterile neutrino. However, in general there is no known constraints on the number and mass scale of right-handed neutrinos. It is assumed that the Majorana neutrinos with heavy mass should be close to the GUT scale. In fact, experiments seem to hint at the possible existence of a fourth type of neutrino which is called a sterile neutrino which would interact even more rarely than the others. If neutrinos are their own antiparticles, they could have played a role in the early universe in expanding period.

Indeed, neutrino physics is one of the most curious and hotly debated topics in particle physics and cosmology, and we are faced with many unsolved problems yet. Without understanding the mystery behind neutrinos, it seems impossible to understand their importance and place in particle physics and cosmology. At this point,  we return to the main idea in the first paragraph, the periodic oscillation hypothesis of the neutrinos.

The main question in our study is this: If neutrinos travel through deformed space-time, does the oscillation period change? Pontecorvo's theory of neutrino oscillation is based on Minkowski space-time. Neutrinos transform into each other due to transitions between mass and flavor eigenstates. However, this transition is not instantaneous, but takes place at certain distances. Oscillation probabilities are defined for a certain critical distance. The oscillation probability and amplitude are formalized in the Minkowski space-time framework and the transition between eigenstates have the Markovian character. But we know from fractional dynamics that such probabilities at the deformed space-times vary with the degree of deformation. There are countless examples of this in physics. Space-time can be deformed by  gravitational perturbations. In this case, it can be expected that the oscillation dynamics deviates from Markovian to the non-Markovian since the deformed space-time leads to the memory effects in the jumping processes between the mass and flavor eigenstates.

The remain organization of the study is as follows: Before the fractional analysis of the neutrino oscillation, firstly, in Section II, we briefly summarize the solution of time evolution of the Dirac equation and give the neutrino transition probability between different flavors. Then, in Section III, we will obtain the solution of the fractional equations and we will introduce a new transition probability which includes deformation of the space-time. Then we finalize the study with discussion and conclusion in Section IV. 

\section{ Theory of the neutrino oscillation}  

We summarize the mathematical background of the neutrino oscillation in the Minkowski space-time following the method in Ref.\,\cite{Mandal2021}. Dirac field theory is defined by introducing the following action
\begin{equation} \label{action}
  S = \int \psi (x) (i\gamma^{\mu} \partial_{\mu} -m ) \psi (t,x)
\end{equation}
which yields the following equation of motion
\begin{equation} \label{Dirac}
	(i\gamma^{\mu} \partial_{\mu} -m ) \psi (t,x) = 0
\end{equation}
where $\gamma^{\mu}$ are the Dirac-gamma matrices $(\mu=0,1,2,3)$, $m$ is the mass of the particle and $\psi$ is the wave function of the spin-1/2 particle. 

The flavor eigenstates of the neutrino denoted by $|\nu_{\alpha} \rangle$ can be represented as linear superposition of the mass eigenstates $|\nu_{j} \rangle $
\begin{equation} \label{neutrino-state}
	|\nu_{\alpha} \rangle = \sum_{j} U_{\alpha j} |\nu_{ j} \rangle, \quad \alpha=e,\mu,\tau, \quad j=1,2,3
\end{equation}
where $U$ is a unitary and the non-diagonal mixing matrix which specifies the composition of each neutrino flavor state. $U$ is given  like Kobyashi-Maskawa as 
\begin{equation} \label{U-matix}
U=\begin{bmatrix}
c_{1}       & s_{1} c_{3}  & s_{1} c_{3}  \\
-s_{1} c_{2}       & c_{1} c_{2}c_{3}-s_{2} s_{3}e^{i\delta}  & c_{1} c_{2}s_{3}-s_{2} c_{3} e^{i\delta}  \\
-s_{1} s_{2}       & c_{1} s_{2}s_{3}-s_{2} s_{3}e^{i\delta}  & c_{1} c_{2}s_{3}-c_{2} c_{3} e^{i\delta} 
\end{bmatrix}
\end{equation}
where $c_{i}=\cos \theta_{i}$, $s_{i}=\sin \theta_{i}$ and $\delta$ is the Dirac CP violating phase.

Conversely, by inverting the relation we obtain the massive neutrino state as $	|\nu_{j}\rangle = \sum_{j} U_{\alpha j} |\nu_{\alpha}  \rangle$. The massive neutrino states are orthonormal as $\langle \nu_{j}|\nu_{k}\rangle=\delta_{jk}$. The mixing matrix has to be unitary transformation as $U^{\dag}U=UU^{\dag}=1$ which guarantees that $\sum_{j}U_{\alpha j}^{*}U_{\beta j}=\delta_{\alpha \beta}$
and $\sum_{\alpha}U_{\alpha j}^{*}U_{\alpha k}=\delta_{jk}$ implies that the flavor states are orthonormal as well, as was expected. 

According to the standard approach, the propagation of mass eigenstate $|\nu_{j}\rangle$ evaluates with time and it can be obtained by the solution of Dirac equation in Eq.\,\ref{Dirac}. The time evolution of the Dirac equation is given by
\begin{equation} \label{Dirac-time}
	i \hbar  \frac{\partial}{\partial t}  \psi_{j} (t, \mathbf{x})  = - \sqrt{\mathbf{p}^{2} c^{2} + m^{2} c^{4} } \frac{ \mathbf{\sigma} . \mathbf{p} }{| \mathbf{p} |} \psi_{j} (t, \mathbf{x}) 
\end{equation}
where $\psi_{j} (t, \mathbf{x}) = \langle \mathbf{x}|v_{j} \rangle$ and $\mathbf{\sigma}$ are Pauli matrices. In the ultra-relativistic limit, the Dirac equation \ref{Dirac-time} can be given as
\begin{equation} \label{Dirac-limit}
	i \hbar  \frac{\partial}{\partial t}  \psi_{j} (t, \mathbf{x})  = -  \left[   |\mathbf{p}| c + \frac{m^{2}_{j} c^{3}}{2|\mathbf{p}|}  \right]    \frac{ \mathbf{\sigma} . \mathbf{p} }{| \mathbf{p} |} \psi_{j} (t, \mathbf{x}) 
\end{equation}
The solution of Eq.\,\ref{Dirac-limit} can be obtained as 
\begin{equation} \label{sol-Dirac}
 \psi_{j} (t, \mathbf{x})  = e^{-i \frac{m^{2}_{j} c^{3}}{2 \hbar |\mathbf{p}|}  t}  e^{-i\frac{pc}{\hbar} t}
\end{equation}
which is equivalent to
\begin{equation} \label{new-state}
	|\nu_{j} (t) \rangle =  e^{-i \frac{m^{2}_{j} c^{3}}{2 \hbar |\mathbf{p}|}  t}  e^{-i\frac{pc}{\hbar} t}   |\nu_{j}  \rangle 
\end{equation}
As a consequence of Eq.\,\ref{neutrino-state}, we obtain
\begin{equation} \label{new-state-new}
	|\nu_{\alpha} (t) \rangle =  \sum_{j} U_{\alpha j} e^{-i \frac{m^{2}_{j} c^{3}}{2 \hbar p}  t}  e^{-i\frac{pc}{\hbar} t}   |\nu_{j}  \rangle 
\end{equation}
The probability amplitude of the initial flavor eigenstate $|\nu_{\alpha} \rangle$ as another flavor eigenstate $|\nu_{\beta}  \rangle$ is given by
\begin{equation} \label{amplitute}
  \langle \nu_{\beta}	|\nu_{\alpha} (t) \rangle =  \sum_{j} U_{\alpha j}  U^{*}_{\beta j}    e^{-i \frac{m^{2}_{j} c^{3}}{2 \hbar p}  t}  e^{-i\frac{pc}{\hbar} t}  
\end{equation}
Hence, the probability for a transition $|\nu_{\alpha} \rangle \rightarrow |\nu_{\beta}  \rangle$ under time evolution is 
\begin{equation} \label{probability-transition}
	\mathcal{P}_{\nu_{\alpha} \rightarrow \nu_{\beta} } (t) =  \sum_{jk}   U_{\alpha j}  U^{*}_{\beta j}  U^{*}_{\alpha k}  U_{\beta k}   e^{-i \frac{( m^{2}_{j} - m^{2}_{k} ) c^{3}}{2 \hbar p}  t}
\end{equation}
where we will set $t=L$. 
	\begin{figure}[h]
	\centering
	\includegraphics[height=4cm,width=7.5cm]{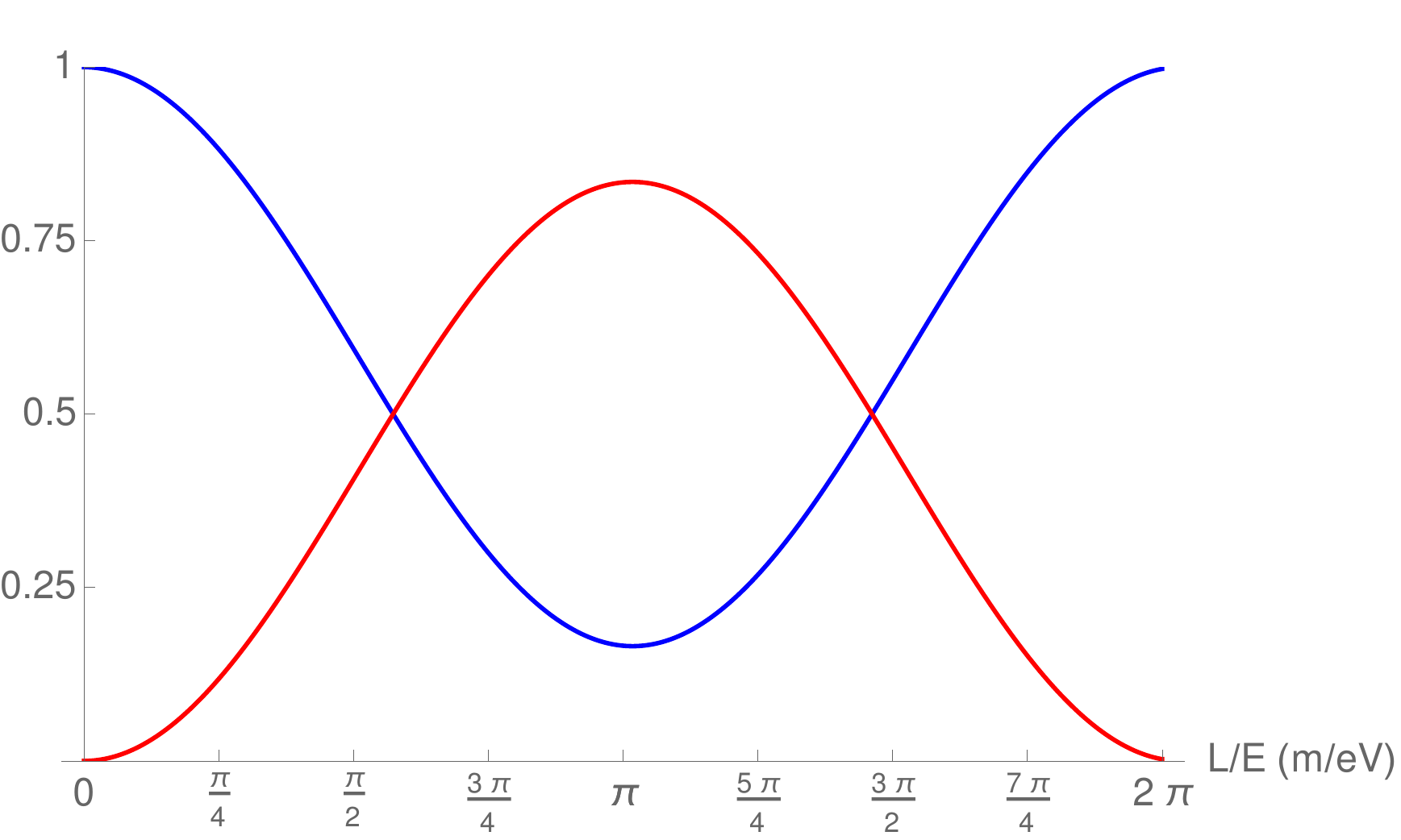}
		\caption{Oscillation probability of electron and muon neutrinos. The red line denotes survival probability electron neutrino while blue line indicates appearance probability of muon neutrino. In this figure, the horizontal axis denotes length/energy i.e. $L/E$ where $L$ and $E$ are the length and energy of the neutrino, respectively }
	\label{Oscillation}
\end{figure}
To see oscillation behavior between two different flavor for instance electron and muon neutrinos we numerically plotted  Eq.\,(\ref{probability-transition}) in Fig.\,(\ref{Oscillation}). In the numerical procedure we set $\theta_{12}=33^{0}$, $\Delta m^{2}_{12}=7.37\times 10^{-5}$ eV$^{2}$ and  $c=3\times10^{8}$ km/s. In Fig.\,(\ref{Oscillation}) red line denotes the transition from electron neutrino to muon neutrino, and the blue line represent the transition from muon neutrino to electron neutrino. This oscillation solution appears completely uniform Minkowski space time and compatible with the standard model of particle physics.

One can see that transition appears at $\pi$ in the radial normalization axis which we will call as transition point $T_{p}$. As it can be seen from Fig.\,(\ref{Oscillation})  the probability of life of the electron neutrino decreases, while the probability of life of the muon neutrino increases. This indicates the conversion of electron neutrino to muon neutrino. It is seen that almost nowhere the oscillation and survival probabilities take the values of zero or one. According to neutrino oscillation experiments, there is always the possibility of oscillation, and therefore there is no situation where the probability of oscillation is equal to zero and the probability of survival is equal to one. This oscillation is repeated continuously along the path.

On the other hand, in this schema, the oscillation length is given by
\begin{equation} \label{o-length}
L^{osc} = 2.47 \frac{E[MeV]}{\Delta m^{2} [eV^{2}] }
\end{equation}
which gives a distance for a complete oscillation which corresponds to an important constraint on the measurement conditions. This constraint implies that the neutrino oscillation can be measured at the $L\sim L^{osc}$. It is supposed that there are no oscillations for $L<<L^{osc} $ and the oscillations are averaged out for $L>>L^{osc}$ due to natural uncertainties of the neutrino energy. The oscillation length is a constant in this conventional schema which says that the energy and mass difference is constant for the flavors.


\section{Anomalous cyclic in the deformed Space-time}

To understand  the deformed space-time on the neutrino oscillation, firstly, it is necessary to accept that particle dynamics depends on the space-time metric. Indeed, the particle motion and oscillation at the microscopic level  in both Minkowski and deformed Minkowski space-time are dependent on space-time since the quantum systems are not perfectly isolated. This fact may be ignored when the dynamic is Markovian. However, if space-time is deformed, the physics completely changes and it turns out that the motion is coupled to the space-time where all perturbation and measurement process leads to decoherence and  deformation of the Hilbert space \cite{Costa2019}. 

We argued above that space-time can be deformed, for instance, by gravitational perturbations. However, we will not try to establish a relationship between deformation and neutrino oscillation using gravitational theories or quantum theories. For simplicity, by using classical stochastic theory we suggest that deformation of the space-time leads divergent characteristic waiting time $T=\int_{0}^{\infty}dt w(t)t$  while the jump length variance $\Sigma^{2}=\int_{-\infty}^{\infty} dx \lambda(x)x^{2}$ is finite where $w(t)=\int_{-\infty}^{\infty} dx \psi(x,t)$ and $\lambda(x)=\int_{0}^{\infty} dt \psi(x,t)$  are waiting time and the jump length probability distribution for $\psi(x,t)$  of particle wave function \cite{Metzler2000,Metzler2001}. The non-Markovian motion has memory and leads fractional dynamics \cite{Metzler2000,Metzler2001,Sokolov2002,Hilfer2000}. In this study, we will analyze the time evolution of the Dirac equation in the fractional framework to obtain the oscillation probability of the neutrinos. For simplicity, we consider the two-neutrino mixing in the vacuum by neglecting three-mixing and the possible existence of the additional sterile neutrino fields.

For the deformed Minkowski space-time the Dirac equation can be written in the fractional form  \cite{Laskin2000,Naber2004,Raspini2001} as  
\begin{equation} \label{f-Dirac}
		i \hbar   D_{t} ^{\eta}  \psi_{j} (t, \mathbf{x})  = -  \left[   |\mathbf{p}| c + \frac{m^{2}_{j} c^{3}}{2|\mathbf{p}|}  \right]    \frac{ \mathbf{\sigma} . \mathbf{p} }{| \mathbf{p} |} \psi_{j} (t, \mathbf{x}) 
	\end{equation}
where $D_{t} ^{\eta} $ denotes the Caputo fractional derivative operator of order $\eta$ and $0<\eta<1$. Here with the aim of retaining the dimensional coherence on both sides of Eq.\,(\ref{f-Dirac}), the energy eigenvalue $E$ varies with power $\eta$. For $\eta=1$, the fractional Schrodinger equation reduces to the standard one. 
The solution of Eq.\,(\ref{f-Dirac}) can be given as
\begin{equation} \label{sol-Dirac-y}
	\psi_{j} (t, \mathbf{x})  =  \chi^{(0)}_{j} (t, \mathbf{x}) \psi^{(0)}_{j}(t,\mathbf{x})
\end{equation}
where two wave function satisfies
\begin{equation} \label{f-Dirac-sol1}
	i   D_{t} ^{\eta} \chi^{(0)}_{j}(t,\mathbf{x})   = -   \lambda^{\eta}_{1} \chi^{(0)}_{j}(t,\mathbf{x}), \quad \lambda_{1}= \frac{ c \mathbf{\sigma} . \mathbf{p} }{\hbar} 
\end{equation}
and 
\begin{equation} \label{f-Dirac-sol2}
	i   D_{t} ^{\eta} \psi^{(0)}_{j}(t,\mathbf{x})   = -   \lambda^{\eta}_{2}  \psi^{(0)}_{j}(t,\mathbf{x}), \quad \lambda_{2} =  \frac{m^{2}_{j} c^{3}}{2 \hbar |\mathbf{p}|}
\end{equation}
In the Riemann-Liouville formalism, the fractional integral of order $\alpha$ is given by the definition \cite{Metzler2000,Metzler2001,Sokolov2002,Hilfer2000}
\begin{equation} \label{RL}
	D_{t}^{\eta} * f(t) = \frac{1}{\Gamma(\eta)} \int_{0}^{t} \left(t-\tau \right)^{\eta-1} f (\tau) d\tau, \quad t>0
\end{equation}
where $\eta$ is any positive number and $\Gamma(\eta)$ is Gamma function. 

The solution of Eq.\,(\ref{f-Dirac-sol1}) and (\ref{f-Dirac-sol2}) can be obtained by taking Laplace transform.  If we perform Laplace transformation on  Eq.\,(\ref{f-Dirac-sol1}) 
\begin{equation} \label{Laplace-t1}
	i   \mathcal{L} \{  D_{t} ^{\eta}  \psi^{(0)}_{j}(t,\mathbf{x}) \}= \mathcal{L} \{ \lambda^ {\eta}_{1,2} \psi^{(0)}_{j}(t,\mathbf{x}) \}
\end{equation}
which yields
\begin{equation} \label{Laplace-t2}
	i   s^{\eta}  \tilde{\psi}^{(0)}_{j}(t,\mathbf{x})  - i s^{\eta-1}  \tilde{\psi}^{(0)}_{j}(0) = \lambda^{\eta}_{1,2}   \tilde{\psi}^{(0)}_{j}(t,\mathbf{x})
\end{equation}
where $ \tilde{\psi}^{(0)}_{j}(t,\mathbf{x})$ is the Laplace transform of the $| \nu_{k} (t) \rangle $, and it can be obtained as
\begin{equation} \label{Laplace-t3}
   \tilde{\psi}^{(0)}_{j}(t,\mathbf{x}) =  \tilde{\psi}^{(0)}_{j}(0)  \frac{i s^{\eta-1} }{is^{\eta}  - \lambda^{\eta}_{1,2}  } 
\end{equation}
By employing inverse Laplace transform one can obtain two independent solutions as  
\begin{equation} \label{sol-12a}
	\chi^{(0)}_{j}(t,\mathbf{x}) =   \chi^{(0)}_{j}(0) E_{\eta}  (-i \lambda^{\eta}_{1}  t^{\eta} )  
\end{equation}
and
\begin{equation} \label{sol-12b}
  \psi^{(0)}_{j}(t,\mathbf{x}) =   \psi^{(0)}_{j}(0) E_{\eta}  (-i \lambda^{\eta}_{2}  t^{\eta} )  
\end{equation}
where $E$ denotes the Mittag-Leffler function \cite{Oldham1974}. It should be noted that the Mittag-Leffler function can be considered as a generalization of the natural exponential one.

Combing Eqs.\,(\ref{sol-12a}) and (\ref{sol-12b}) into Eq.\,(\ref{sol-Dirac-y}), general solution of fractional Dirac equation can be written as
\begin{equation} \label{Sol1-Dirac}
	\psi_{j}(t,\mathbf{x}) =   \psi^{(0)}_{j}(0) E_{\eta}  (-i \lambda^{\eta}_{1}  t^{\eta} )  E_{\eta}  (-i \lambda^{\eta}_{2}  t^{\eta} )  
\end{equation}
As a consequence of Eq.\,(\ref{neutrino-state}), we obtain
\begin{equation} \label{frac-state}
	| \nu_{j} (t) \rangle   =    \sum_{j=1}^{3} U_{\alpha j}    E_{\eta}  (-i \lambda^ {\eta}_{1}  t^{\eta} )  E_{\eta}  (-i \lambda^ {\eta}_{2}  t^{\eta} )  | \nu_{j}  \rangle
\end{equation}
The series expression of the Mittag-Leffler function is given as 
\begin{equation} \label{Mittag-Leffler}
    E_{\eta}  (-i \lambda^ {\eta}  t^{\eta} )  =  \sum_{j=0}^{\infty} \frac{(-i \lambda^ {\eta} t^{\eta})^{j}  }{\Gamma(1+\eta j)}
\end{equation}
For $\eta=1$, Eq.\,(\ref{Mittag-Leffler}) reduces to the standard exponential form, whereas for $0<\eta < 1$, initial stretched exponential behaviour
\begin{equation} \label{KWW}
	E_{\eta}  (-i \lambda^ {\eta}  t^{\eta} )  \sim   \exp \left( - \frac{i \lambda^ {\eta} t^{\eta} }{\Gamma(1+\eta )}  \right)
\end{equation}
turns over to the power-law long-time behavior
\begin{equation} \label{Power}
	E_{\eta}  (-i \lambda^ {\eta}  t^{\eta} )  \sim  \frac{1 }{i \Gamma(1-\eta)\lambda^ {\eta} t^{\eta}} .
\end{equation}
By using the above expression, we can give Eq.\,(\ref{frac-state}) in the simplified form
\begin{equation} \label{frac-state-simlify}
	| \nu_{\alpha} (t) \rangle   =    \sum_{j=1}^{3} U_{\alpha j}    e^{-i (\frac{m^{2}_{j} c^{3}}{2 \hbar p} )^{\eta} t^{\eta}  }  e^{ -i (\frac{pc}{\hbar})^{\eta} t^{\eta}  }   | \nu_{j}  \rangle
\end{equation}
In this case, the probability amplitude becomes
\begin{equation} \label{f-amplitute}
	\langle \nu_{\beta}	|\nu_{\alpha} (t) \rangle =  \sum_{j} U_{\alpha j}  U^{*}_{\beta j}     e^{-i (\frac{m^{2}_{j} c^{3}}{2 \hbar p} )^{\eta} t^{\eta}  }  e^{ -i (\frac{pc}{\hbar})^{\eta} t^{\eta}  }
\end{equation}
Hence, the probability for a transition $| \nu_{k}  \rangle \rightarrow | \nu_{\beta}  \rangle$ under time evolution is
\begin{equation} \label{frac-prob}
	\mathcal{P}_{\nu_{\alpha} \rightarrow \nu_{\beta} } (t) =  \sum_{jk}   U_{\alpha j}  U^{*}_{\beta j}  U^{*}_{\alpha k}  U_{\beta k}  e^{-i (\frac{(m^{2}_{j} - m^{2}_{j} )   c^{3}}{2 \hbar p} )^{\eta} t^{\eta}  }
\end{equation}
where we will set $t=L$. 
One can see that the transition probability has the stretched exponential form which is often called as KWW function \cite{Kohlrausch1847,Williams1970}.
It should be noted that for $\eta=1$, Eq.\,(\ref{frac-prob}) reduces to Eq.\,(\ref{probability-transition}).

\begin{figure}[h!]
	\begin{subfigure}
		\centering
		\includegraphics[height=3.4cm,width=7cm]{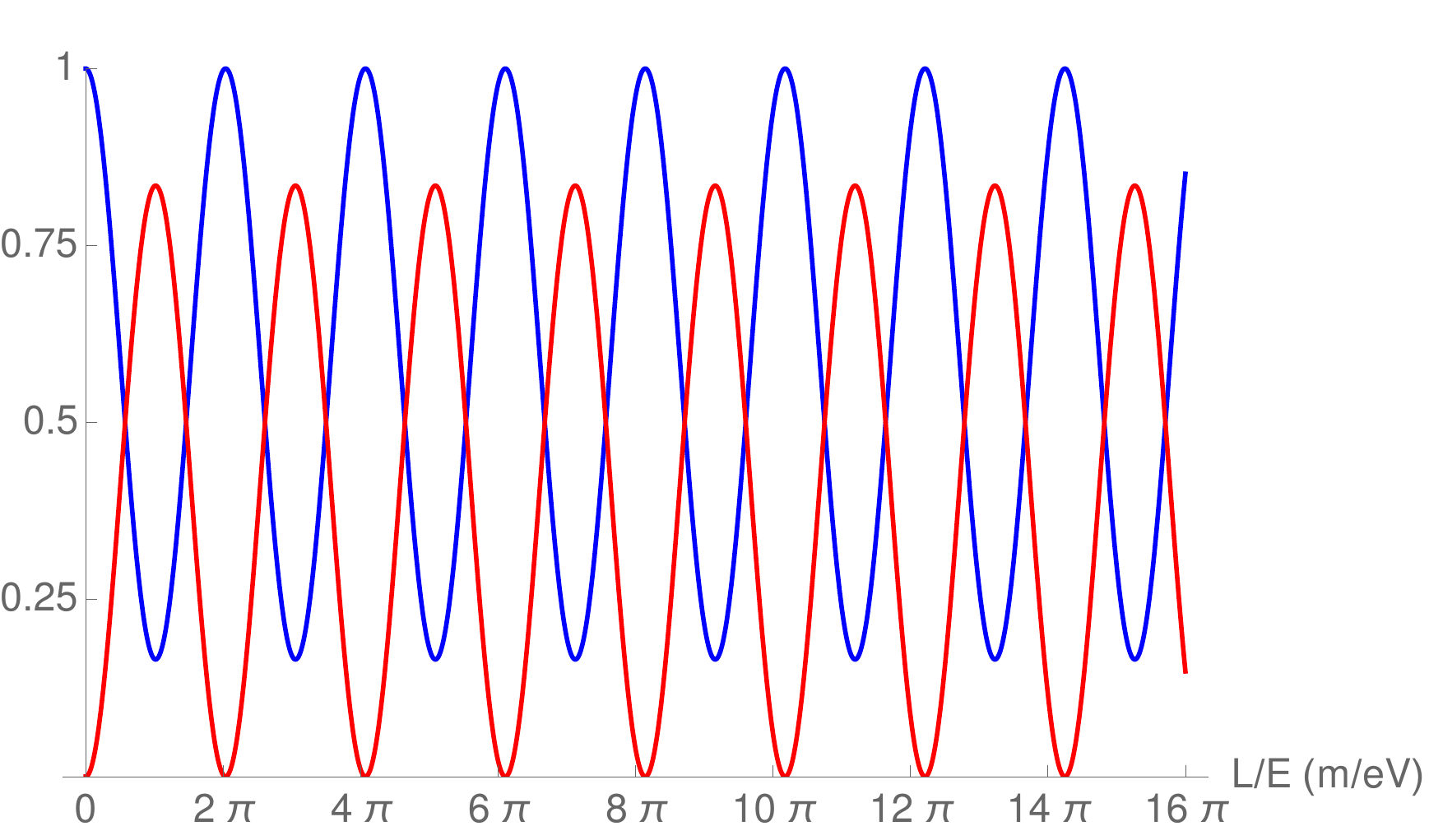}
		\label{sfig:testa}
	\end{subfigure}
	\begin{subfigure}
		\centering
		\includegraphics[height=3.4cm,width=7cm]{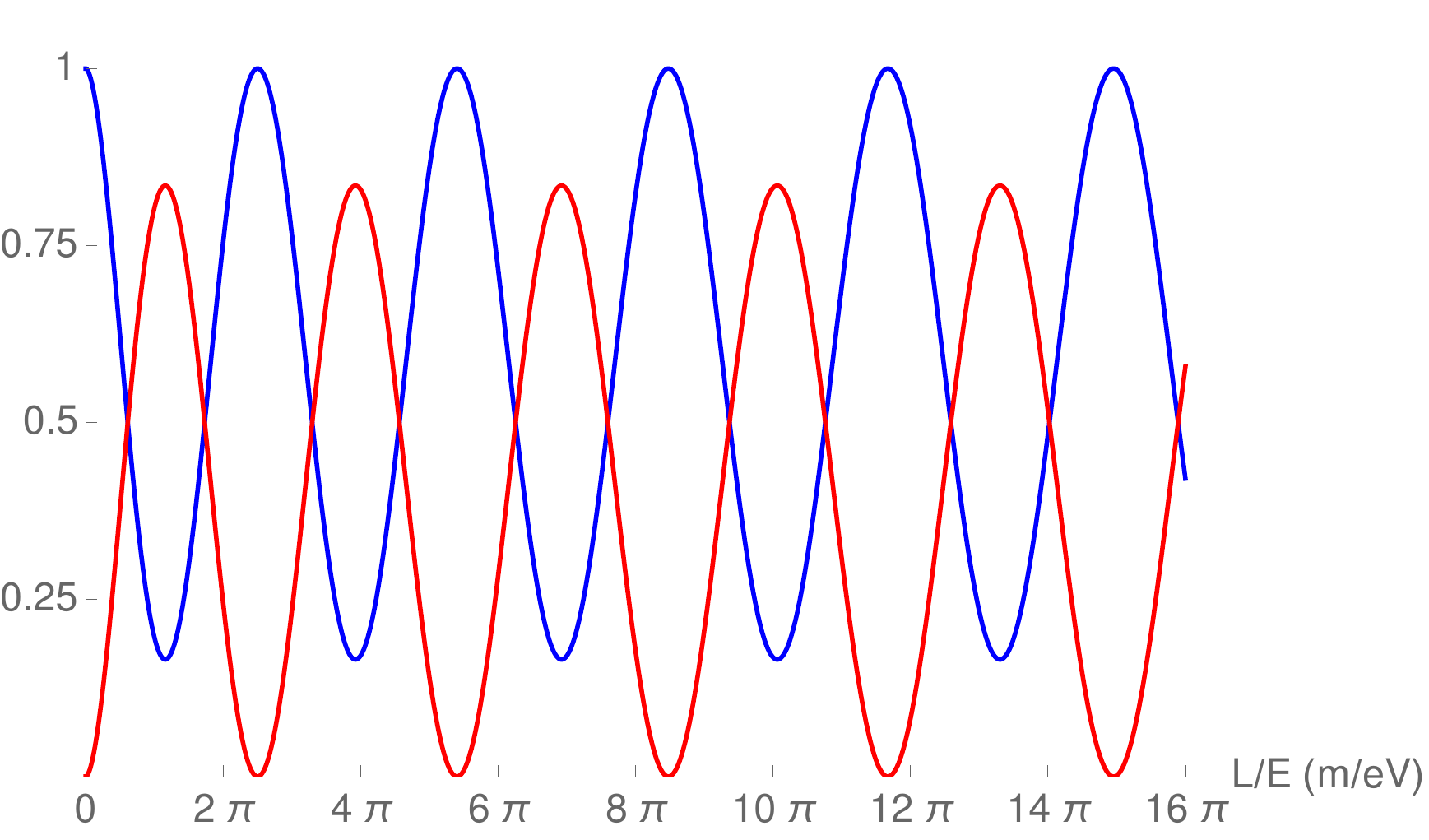}
		\label{sfig:testb}
	\end{subfigure}
	\begin{subfigure}
		\centering
		\includegraphics[height=3.4cm,width=7cm]{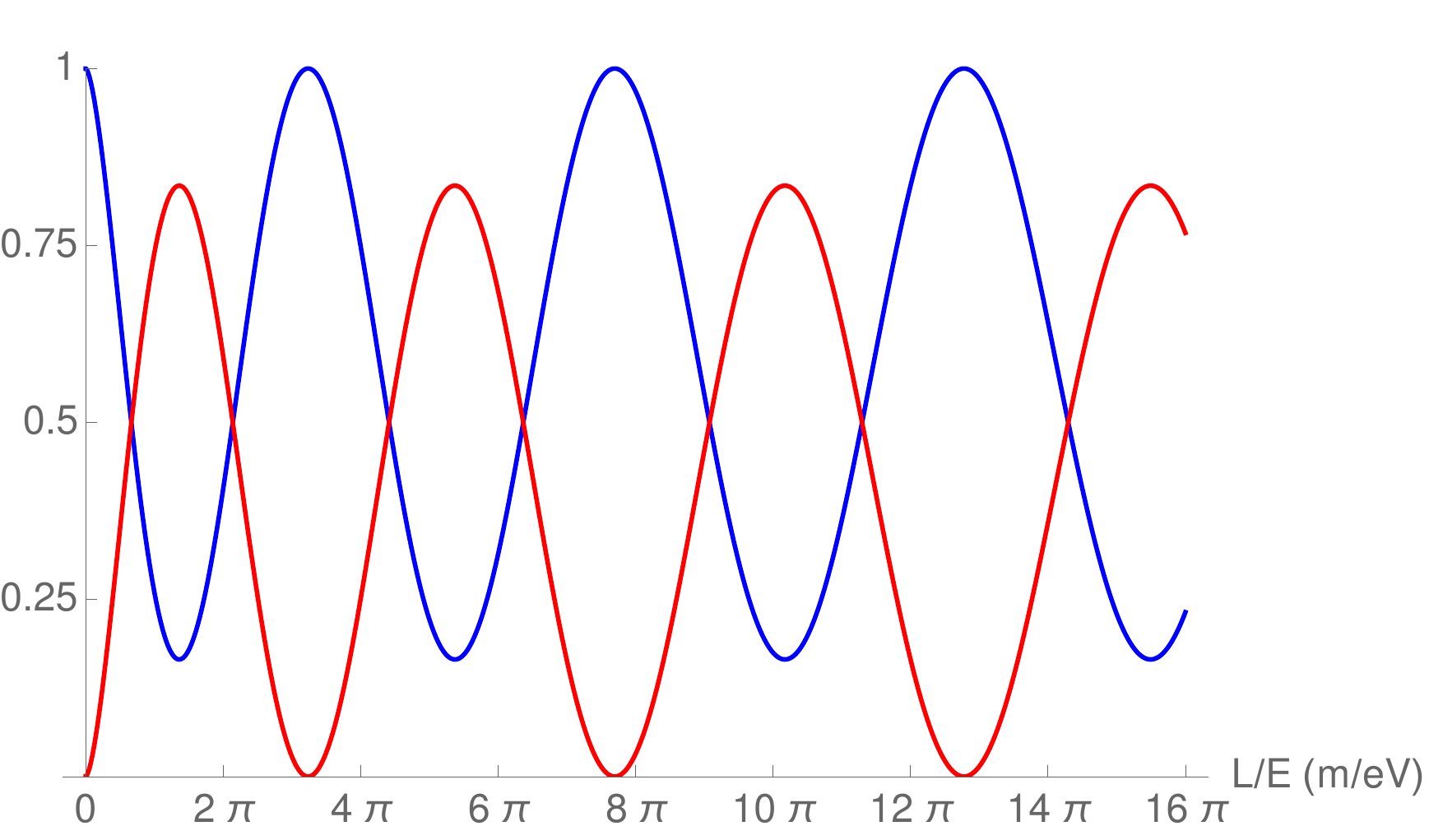}
		\label{sfig:testa}
	\end{subfigure}
	\begin{subfigure}
		\centering
		\includegraphics[height=3.4cm,width=7cm]{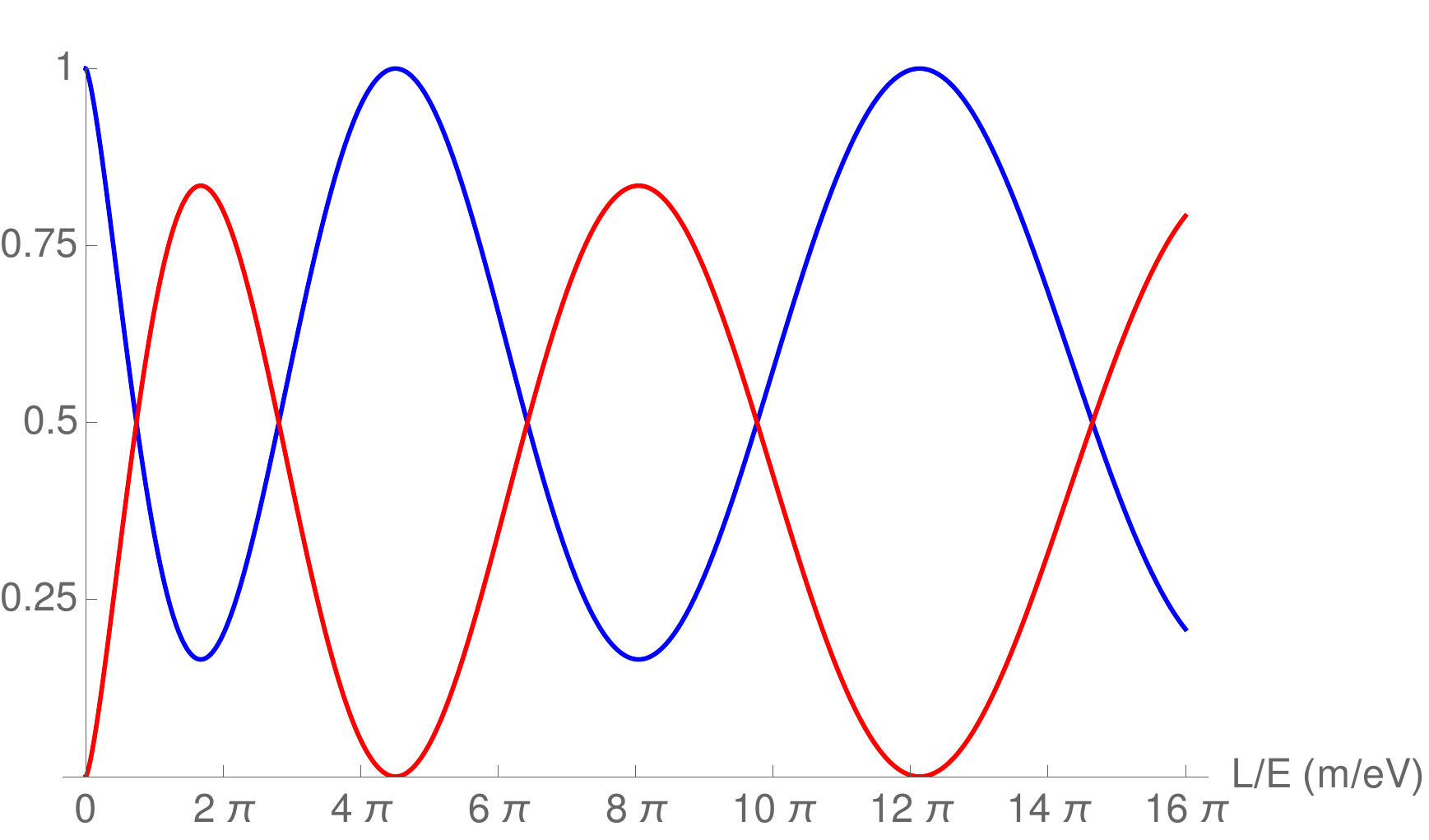}
		\label{sfig:testa}
	\end{subfigure}
	\caption{The oscillation probabilities of electron and muon neutrinos. The red line denotes electron neutrino while the blue line indicates muon neutrino. In (a) the oscillation probability in the Minkowski space-time for $\eta=1$ which indicates the ordinary Minkowski space-time. The oscillation probabilities in the deformed Minkowski space-time are given in (b)-(d) subfigures. The deformation parameters of the Minkowski space-time are $\eta=0.9$ in (b) $\eta=0.8$ in (c) and $\eta=0.7$ in (d). In this figure, the horizontal axis denotes length/energy i.e. $L/E$ where $L$ and $E$ are the length and energy of the neutrino, respectively.}
	\label{Fractional-os}
\end{figure}
We analytically obtained neutrino oscillation probability for the deformed Minkowski space-time above. In order to see the effect of the fractional order on the oscillation probability we numerically solved Eq.\,(\ref{frac-prob}) for various $\eta$ and plotted all numerical results in Fig.\,(\ref{Fractional-os}). In the numerical procedure we set $\theta_{12}=33^{0}$, $\Delta m^{2}_{12}=7.37\times 10^{-5}$ eV$^{2}$ and  $c=3\times10^{8}$ km/s. 

The oscillation probabilities of electron and muon neutrinos ar given in Fig.\,(\ref{Fractional-os}). The red line denotes the electron neutrino while the blue line indicates the muon neutrino. In Fig.\,(\ref{Fractional-os})(a) oscillation probability in the Minkowski space-time for $\eta=1$ which indicates ordinary Minkowski space-time. The oscillation probabilities in the fractional Minkowski space-time are given in Fig.\,(\ref{Fractional-os})(b)-(d) subfigures. The fractional parameters of Minkowski space-time are set as $\eta=0.9$ in (b) $\eta=0.8$ in (c) and $\eta=0.7$ in (d). In this figure, the horizontal axis denotes length/energy i.e. $L/E$ where $L$ and $E$ are the length and energy of the neutrino, respectively.

As can be seen from Fig.\,\ref{Fractional-os} the probability of life of the electron neutrino decreases, while the probability of life of the muon neutrino increases as well in Fig.\,\ref{Oscillation}. This oscillation points out the transition from electron neutrino to muon neutrino. Additionally, the oscillation and survival probabilities remain the same without changing.  For example, for $\eta =1.0$ which corresponds to the Minkowski space-time where the oscillations are quite smooth and appear subsequently at the transition points $T_{p} = \pi, 3\pi, 5\pi,...$ and so on as seen from Fig.\,\ref{Fractional-os} (a) or Table.\ref{tab1}. 

However,  the neutrino oscillation probability for the deformed Minkowski space-time is quite different as can be seen from Fig.\,\ref{Fractional-os}  (b)-(d) unlike for $\eta=1$ as well as in Fig.\,\ref{Oscillation} or Fig.\,\ref{Fractional-os}(b). However, for $\eta < 1$ i.e neutrino oscillation in the deformed Minkowski space leads to two main and important results: 

Firstly, as it can be seen from  Fig.\,\ref{Fractional-os}(a) or Table.\ref{tab1}  the oscillation transition points takes  the values $T_{p} = \pi, 3\pi, 5\pi,...$ for $\eta =1$.  However, for $\eta < 1$,  the transition points slide to the right and these increments regularly increase for the subsequent oscillations as seen in  Fig.\,\ref{Fractional-os} (b)-(d). For $\eta < 1$, the amounts of the increments in the transition points are given in Table.\ref{tab1}. One can see that the shifts in the transition points increase when $\eta$ values decrease. These results show the presence of anomalous behavior in the neutrino oscillation \footnote{I would like to bring to the attention of the reader that the subject of phase shift in neutrino oscillation was first considered in Ahluwalia and Burgard's articles \cite{Ahluwalia1996,Ahluwalia1998}.}.

Secondly, the period length of the oscillation, i.e., oscillation time, regularly increases for each $\eta$ values depending on the shifts of the transition points  \footnote{This behavior may be interpreted as gravitational red and blue shifts.}. This behavior can also be seen from  Fig.\,\ref{Fractional-os}  (b)-(d). One can see that the period lengths dramatically increase when $\eta$ decreases. The increments in the period length are given Table.\ref{tab2}.  These results clearly indicate that deformation in the space-time extends the survival probabilities of flavors.
On the other hand, we expect that if a particle whose period length changes in a deformed Minkowski space-time continues to move in a homogeneous Minkowski space-time, it is expected that the particle will return to its original period-length state and can continue to oscillate in its original form.
	
\begin{table}[h!]
		\caption{Oscillation points for various $\eta$ values.}
	\begin{tabular}{lllll}
		\toprule
	$\eta$ \quad \quad &	$T_{p1}$  \quad \quad  & $T_{p2}$ \quad  \quad &$T_{p3}$ \quad \quad & $T_{p4}$  \\
		\colrule
	$\eta=1.0 $  \quad \quad  &  $\pi$  \quad  \quad &$3\pi$  \quad \quad  &$5\pi$  \quad  \quad& $7\pi$\\
	$\eta=0.9$ \quad  \quad &$1.11\pi$  \quad  \quad &$3.88\pi$ \quad \quad &$6.90\pi$  \quad \quad & $0.02\pi$\\
	$\eta=0.8$ \quad  \quad &$1.38\pi$ \quad  \quad  &$5.34\pi$ \quad \quad  &$10.17\pi$ \quad   \quad & $15.47\pi$\\
	$\eta=0.7$ \quad  \quad &$1.67\pi$ \quad \quad  &$8.05\pi$  \quad  \quad &$16.62\pi$ \quad \quad  & $27.04\pi$\\
		\botrule
	\end{tabular}
	\label{tab1}
\end{table}

\begin{table}[h!]
		\caption{The period lengths of the  oscillation for various $\eta$ values.}
	\begin{tabular}{lllll}
		\toprule
		$\eta$ \quad \quad& 	$l_{1}$  \quad \quad &  $l_{2}$ \quad \quad &  $l_{3}$ \quad \quad  &  $l_{4}$  \\
		\colrule
		$\eta=1.0$ \quad \quad& $2\pi$ \quad \quad & $2\pi$ \quad \quad & $2\pi$ \quad \quad &  $2\pi$\\
		$\eta=0.9$ \quad \quad& $2.50\pi$ \quad \quad & $2.88\pi$ \quad \quad &  $3.04\pi$ \quad \quad &  $3.19\pi$\\
		$\eta=0.8$ \quad \quad& $3.22\pi$ \quad \quad & $4.49\pi$ \quad \quad &  $5.14\pi$ \quad \quad & $5.47\pi$\\
		$\eta=0.7$ \quad \quad&$4.52\pi$ \quad \quad & $7.59\pi$ \quad \quad  &  $9.50\pi$ \quad \quad & $10.95\pi$\\
		\botrule
	\end{tabular}
	\label{tab2}
\end{table}


\section{Discussion and Conclusion}

In this study we briefly summarized the standard theory of the neutrino oscillation for the Minkowski space-time and gave the oscillation picture in Fig.\,\ref{Oscillation}. Then, we obtained a new oscillation probability function of the neutrinos for the deformed Minkowski space-time. We show that anomalous cyclic behavior in the neutrino oscillation can appear in the deformed Minkowski space-time, which cannot be explained by the standard theory of the particle physics and can be ignored in the experimental studies without being theoretically carried out.

We showed that when space-time is deformed, for instance, by gravitational waves, neutrino oscillation transition time and its period length gradually increase between subsequent events as in Fig.\,\ref{Fractional-os}  (b)-(d) and Table.\ref{tab2}. Therefore, this anomalous behavior violates the oscillation constraint  of the measurement condition $L\sim L^{osc}$ in the standard approach  
Our results provide that  for $L<<L^{osc} $ there are no oscillations, however, for $L>>L^{osc}$ the oscillations can appear in the deformed Minkowski space-time.  For $L>>L^{osc}$, the change in neutrino energy and the mass difference is not due to inherent uncertainties unlike as well conventional picture. Therefore, the oscillations should not be averaged. 

According to our results, different oscillation lengths clearly indicate a deeper physics. These significant results  are reported in this study for the first time. These results have the potential to explain why different detectors account to the different numbers of the electron neutrinos in the experimental studies. Furthermore, these results reveal that an increase of the period length cause a loss in the masses of neutrinos in the deformed space-time that can lead to uncertainties in the measurements of the mass. Some important questions for example mass limits, mass hierarchy, mass gain mechanism and measurements of different numbers of electrons in different detectors can appear due to the anomalous oscillations of the neutrino which can be observed experimentally. Furthermore, looking at the problem in reverse, if an observational limit can be obtained in experiments for the anomalous oscillation, it may be possible to determine a more accurate range for the neutrino mass and also to determine the extent to which gravitational waves deform space-time. All the results obtained here can be generalized to the Friedman-Robertson-Walker space-time. 

Understanding the mystery behind neutrino physics will obviously play an important role in cosmology and in particle physics. Therefore, the observation of such an anomaly in oscillation could lead to a new era in neutrino physics. As a result, although it is very difficult to detect the cyclic anomalies in the neutrino oscillations, it may not be impossible to detect them \footnote{I would like to point out that one method of the testing the increasing phase shift in neutrino oscillation may be to look for a change in the measured neutrino flux on Earth at the moment when the gravitational waves are observed after the black holes, neutron star collisions or supernova explosions. Observation of a correlation between rippling gravitational waves and neutrino flux may be indirect evidence for phase shifting in neutrino oscillation.}. The detection of these cyclic anomalous can lead to a new discussion in neutrino physics.

\section{Update} 

The relic neutrinos are the most important problem as far as  the Cosmic Microwave Background (CMB). As it is known, the CMB is a remnant that appears at the approximately 380,000 years after cosmic beginning, and its temperature was determined to be approximately $T \approx 2.725$ K. It is also assumed that at the cosmic beginning of the Universe, the neutrinos separated from the matter at the $t=1$ s. This is called the cosmic neutrino background (CNB or C$\nu$B). In the theoretical calculations, it is estimated that C$\nu$B has a temperature of approximately $T \approx 1.95$ K. 

Indeed, the direct  \cite{Weinberg1962} and indirect methods  \cite{Follin2015} are proposed to observe the  neutrino relic. However, no successful observation has yet been achieved so far. For example, one of the methods is analyze the images taken from the Planck data. The energies of the cold regions in these data images are estimated to be $T \approx 1.95$ K and it is assumed that these regions may point to the relic neutrinos. In this sense, one can expect that the James Webb telescope can provide new and more detailed information on this background image. On the other hand, the relic neutrinos are tried to be observed in direct particle experiments.
Weinberg proposed a method to directly detect the relic neutrinos \cite{Weinberg1962}. According to this scenario, neutrino can be captured on a tritium target  \cite{Weinberg1962} on the process: $\nu +  {}^3$H $\rightarrow {}^3$He $+ e^{-}$.  To detect relic neutrinos many methods  and experimental setups have been proposed so far for example  KATRIN \cite{Aker2022}, PTOLEMY \cite{Baracchini2018,Betti2019} etc.

At this point, we may ask the question if there is a third way to detect the neutrino relic. In this study, we suggested that anomalous cycles can be observed in neutrino oscillation. Can a relationship be established between possible anomalous cycles in neutrino oscillation and neutrino age? This idea is actually quite interesting. If a relationship can be established between the anomalous cycle period, oscillation length, and the age of neutrinos, a new way to detect neutrino relics could be opened. In order to establish a relationship between oscillation length and aging, theoretical and experimental limits should be determined. We will not be able to do that in this article, but if this work can be done, there may be a way to observe or measure the neutrino relic.

\textbf{Acknowledgments--} I would like to  thank O. Demirkap for the useful private communication. I modified the Mathematica program to plot the oscillation probability in the Mc.S. thesis of O. Demirkap.  I am grateful to Dharam Vir Ahluwalia for the useful comments. I am also grateful to Güngör Gündüz for the English proofreading of this article. I also thank Durmuş Ali Demir, Ali Ulvi Yılmazer, Muhammed Deniz, Burra Sidharth, Özgür Ökcü and Suat Özkorucuklu for useful discussions. This work was partially supported by Istanbul University with Research Project: FYO-2021-38105 which is titled “Investigation of the Cosmic Evolution of the Universe”.

\bibliography{Neutrino}

\end{document}